\documentclass
[%
	final,  
	peerreviewca,
]
{IEEEtran}%
\IEEEoverridecommandlockouts%

\usepackage[
    hidelinks,
]{hyperref}

\usepackage{graphicx}

\usepackage[
    noend,
]{algpseudocode}

\usepackage[
    table,
]{xcolor}

\usepackage{xspace}

\usepackage{cite}

\usepackage{relsize}

\usepackage{			
	amsmath,
	amssymb,
	amsfonts,
}

\usepackage{siunitx}

\usepackage{tikz}
\usetikzlibrary{
	positioning,				
	shapes.geometric,			
	decorations.pathreplacing,	
    decorations.markings,       
	shadows,					
	patterns,					
    backgrounds,                
    calc,                       
}

\usepackage[
	toc=True,
	numberline=False,
	style=tree,
	sort=standard,
	nopostdot=True,
	acronym,
]{glossaries}

\glsdisablehyper  
\setacronymstyle{long-sm-short}  

\newacronym{sota}{SotA}{State of the Art}
\newacronym
    [longplural={Degrees of Freedom}]
    {dof}{DoF}{Degree of Freedom}

\newacronym{3gpp}{3GPP}{3rd Generation Partnership Program}

\newacronym{embb}{eMBB}{Enhanced Mobile Broadband}
\newacronym{urllc}{URLLC}{Ultra Reliable and Low Latency Communications}
\newacronym{mmtc}{mMTC}{Massive Machine Type Communications}

\newacronym{tx}{Tx}{Transmit}
\newacronym{rx}{Rx}{Receive}

\newacronym
    [longplural={Channel State Information at Transmitter}]
    {csit}{CSIT}{Channel State Information at Transmitter}
\newacronym{ber}{BER}{Bit Error Rate}
\newacronym{snr}{SNR}{Signal to Noise Ratio}

\newacronym{sgd}{SGD}{Stochastic Gradient Descent}
\newacronym{mmse}{MMSE}{Minimum Mean Squared Error}
\newacronym{zf}{ZF}{Zero Forcing}

\newacronym{dtft}{DTFT}{Discrete Time Fourier Transform}
\newacronym{dft}{DFT}{Discrete Fourier Transform}
\newacronym{fft}{FFT}{Fast Fourier Transform}

\newacronym{ofdm}{OFDM}{Orthogonal Frequency Division Multiplex}
\newacronym{ofdma}{OFDMA}{Orthogonal Frequency Division Multiple Access}
\newacronym{noma}{NOMA}{Non Orthogonal Multiple Access}

\newacronym{siso}{SISO}{Single Input Single Output}
\newacronym{simo}{SIMO}{Single Input Multiple Output}
\newacronym{miso}{MISO}{Multiple Input Single Output}
\newacronym{mimo}{MIMO}{Multiple Input Multiple Output}

\newacronym{ml}{ML}{Machine Learning}
\newacronym{dl}{DL}{Deep Learning}
\newacronym{rl}{RL}{Reinforcement Learning}

\newacronym{nn}{NN}{Neural Network}
\newacronym{acnn}{AcNN}{Actor Neural Network}
\newacronym{crnn}{CrNN}{Critic Neural Network}

\newacronym{dqn}{DQN}{Deep Q Network}
\newacronym{ddpg}{DDPG}{Deep Deterministic Policy Gradient}
\newacronym{sac}{SAC}{Soft Actor-Critic}
\newacronym{ppo}{PPO}{Proximal Policy Optimization}

\newacronym{ntn}{NTN}{Non Terrestrial Networks}
\newacronym{leo}{LEO}{Low Earth Orbit}
\newacronym{meo}{MEO}{Medium Earth Orbit}
\newacronym{geo}{GEO}{Geostationary}

\newacronym{sic}{SIC}{Successive Interference Cancellation}
\newacronym{los}{LoS}{Line-of-Sight}
\newacronym
    [longplural={Angles of Departure}]
    {aod}{AoD}{Angle of Departure}
\newacronym{awgn}{AWGN}{Additive White Gaussian Noise}
\newacronym{aoa}{AoA}{Angle of Arrival}
\newacronym{ula}{ULA}{Uniform Linear Array}
\newacronym{rsma}{RSMA}{Rate-Splitting Multiple Access}
\newacronym{sdma}{SDMA}{Space-Division Multiple Access}
\newacronym{oma}{OMA}{Orthogonal Multiple Access}
\newacronym{mrt}{MRT}{Maximum Ratio Transmission}
\newacronym{iui}{IUI}{Inter-User Interference}
\newacronym{uav}{UAV}{Unmanned Aerial Vehicle}
\newacronym{slnr}{SLNR}{Signal-to-Leakage-and-Noise Ratio}
\newacronym{sinr}{SINR}{Signal-to-Interference-plus-Noise Ratio}

\newglossaryentry{ex}{%
	name={example},
	description={an example},
}%

\title
{%
	Flexible Robust Beamforming for Multibeam Satellite Downlink using Reinforcement Learning%
	\thanks{This work was partly funded by the German Ministry of Education and Research (BMBF) under grant 16KISK016 (Open6GHub) and 16KIS1028 (MOMENTUM) and the European Space Agency (ESA) under contract number 4000139559/22/UK/AL (AIComS).}%
}

\author{%
	\IEEEauthorblockN{%
		Alea~Schröder,\ 
		Steffen~Gracla,\ 
		Maik~Röper,\ 
		Dirk~Wübben,\ 
		Carsten~Bockelmann,\ 
		Armin~Dekorsy
	}%
	\IEEEauthorblockA{%
		Dept. of Communications Engineering, University of Bremen, Bremen, Germany\\
		{Email: \{schroeder, gracla, roeper, wuebben, bockelmann, dekorsy\}@ant.uni-bremen.de}
	}%
}
\input{99_custommacros.sty}

\begin{document}
	\maketitle%
	%
	

\begin{abstract}
\gls{leo} satellite-to-handheld connections herald a new era in satellite communications. \gls{sdma} precoding is a method that mitigates interference among satellite beams, boosting spectral efficiency. While optimal \gls{sdma} precoding solutions have been proposed for ideal channel knowledge in various scenarios, addressing robust precoding with imperfect channel information has primarily been limited to simplified models.
However, these models might not capture the complexity of LEO satellite applications.
We use the \gls{sac} deep \gls{rl} method to learn robust precoding strategies without the need for explicit insights into the system conditions and imperfections.
Our results show flexibility to adapt to arbitrary system configurations while performing strongly in terms of achievable rate and robustness to disruptive influences compared to analytical benchmark precoders.
\end{abstract}


\glsresetall  
%
	

\section{Introduction}
\label{sec:introduction}
The upcoming sixth generation standard of mobile communications will seek to extend our classic terrestrial networks to so-called 3D networks by integrating communication satellites and \glspl{uav} \cite{3GPP.TR.38.863}. This additional degree of freedom is expected to boost continuous global coverage, improve balancing traffic demand surges, and increase outage protection \cite{Leyva-Mayorga2020, Qu}.
In contrast to higher satellite orbits, the \gls{leo} offers relatively low latency and path loss, and reduced deployment cost. Therefore, \gls{leo} satellites are targeted as a key component of future \gls{ntn}~\cite{3GPP.TR.38.863}.
The satellites' downlink transmit power can be steered by \gls{sdma}~precoding.
The precoder optimizes each users' signal power while mitigating inter-user interference, achieving better spectral efficiency~\cite{vazquez2018precoding}.
Conventional \gls{sdma} precoding techniques, however, are challenged by errors in position estimation caused by outdated \gls{csit}, which can degrade the performance quickly~\cite{liu2022robust}.
Under real circumstances, a variety of perturbing effects may further influence the transmission quality. They are induced by, for example, the high relative velocity of \gls{leo} satellites and atmospheric influences.

In order to design a robust precoder, the authors in~\cite{liu2022robust} maximize the achievable rate, taking imperfect position knowledge into account, and propose a low complexity algorithm using supervised learning. In~\cite{roper2023robust}, an analytical robust precoder is derived that utilizes the second order statistics of the channel to maximize the mean \gls{slnr}. \cite{schroeder2023wsa} investigate \gls{rsma} to deal with errors in position measurements, by splitting the user messages into individually precoded common and private parts.
In our previous paper~\cite{gracla2023satellite}, we take a first look at using deep \gls{rl} for the purpose of beamforming in the presence of imperfect \gls{csit}.
\gls{dl}, as a sub-set of \gls{ml}, infers and iteratively tunes the parameters of a \gls{nn} based on a data set such that the \gls{nn} approximately maximizes an objective.
These data-driven \gls{dl} approaches have achieved noteworthy performance across most domains in the past decade, and their application in communication networks has been under intense study, \eg \cite{dahrouj2021overview}.
\gls{ml} is of particular interest for robust algorithms since input data with uncertainty are often seen as desirable and positive during the learning process~\cite[Chp.~7.5]{goodfellow2017deep}: to prevent over-fitting on the available data, practitioners will frequently add noise on their data as a regularizing measure. 
In~\cite{gracla2023satellite}, we assume a multi-satellite downlink scenario and use \gls{rl} to train a \gls{nn} that takes as input a channel matrix estimate to output a precoding matrix that approximately maximizes the achievable sum rate.
We use the \gls{sac} \gls{rl} algorithm~\cite{haarnoja2019soft} that allows for continuous-valued output and maintains a measure of uncertainty about its decisions.
This measure of uncertainty is used to guide the learning process to be sample efficient.
We demonstrate the feasibility of this approach and its strong robustness to disruptive influences comparing to the common \gls{mmse} precoding approach.

In this paper we move on to a single satellite downlink scenario that is more challenging due to covering much larger user distances.
We compare our performance to the popular \gls{mmse} precoder, as well as the above mentioned analytical robust precoder~\cite{roper2023robust}, both of which strongly favor this spatially decoupled scenario.
In \refsec{sec:setup} and \refsec{sec:conventionalprecoders}, we  first discuss the satellite downlink model and the two benchmark precoders. Afterwards, we outline the learning approach in \refsec{sec:OURAPPROACH} and discuss the adjustments for this more complex scenario. Finally, we present our findings in \refsec{sec:experiments} and conclude in \refsec{sec:conclusions}.

\textit{Notations}: Boldface $\mathbf{x}$ and $\mathbf{X}$ denote vectors resp. matrices. $\mathbf{I}_N$ is an $N \times N$ identity matrix.
We use the operators Transpose~$\{\cdot\}^\text{T}$, Hermitian~$\{\cdot\}^\text{H}$, Expectation~$\mathbb{E}\{\cdot\}$, Hadamard product~$\circ$, absolute value~$|\cdot |$ and Euclidean norm~$\| \cdot \|$.


%
	

\section{Downlink Satellite Communication Model}
\label{sec:setup}

We examine a single-satellite multi-user downlink scenario as depicted in \reffig{fig:geometry}. The \gls{leo} satellite is equipped with a \gls{ula} consisting of $\numantennasper$ antennas with an inter-antenna-distance $\antdist$ and transmit gain $\satgain$. The $\numusers$ users are assumed to be handheld devices with just one receive antenna and low receive gain $\usergain$. The \gls{los} channel $\csivector_\useridx \in \mathbb{C}^{1\times \numantennasper }$ between the satellite and user $\useridx$ is modeled by


\begin{align}
\label{eq:channel_perfect}
    \csivector_{\useridx}(\aod_{\useridx}) = 
     \frac{1}{\sqrt{{\pathloss_\useridx}}}  \text{e}^{-j \kappa_{\useridx} }  \steeringvec_{\useridx} ( \cos(\aod_{\useridx})) \ .
\end{align}
The path loss $\pathloss_\useridx$ is the linear representation of the free space path loss $\freespacepathloss_\useridx$ influenced by large scale fading $\largescalefading_\useridx \sim \mathcal{N}(0, \sigma_{\text{LF}}^2) $ with $\pathloss_\useridx^\text{dB} = \freespacepathloss_\useridx^\text{dB} + \largescalefading_\useridx^\text{dB}$. The linear free space path loss $\freespacepathloss_\useridx$ for a given wavelength $\wavelength$ and a satellite-to-user distance $\dist_\useridx$ is 

\begin{align}
    \freespacepathloss_\useridx = \frac{16\pi^2 \dist^2_{\useridx}}{\wavelength^2{\usergain \satgain }}\ .
\end{align}

The overall phase shift from the satellite to user $k$ corresponds to $\kappa_\useridx \in [0,2\pi]$, while the relative phase shifts from the $\numantennasper$ satellite antennas to user $\useridx$ are determined by the steering vector $\steeringvec_\useridx \in \mathbb{C}^{1 \times \numantennasper}$. The $\antidx$-th entry of the steering vector $\steeringvec_\useridx(\cos(\aod_\useridx))$ calculates as follows

\begin{align}
\label{eq:steering_vec}
     \steeringentry^\antidx_{\useridx}(\cos(\aod_{\useridx})) = \text{e}^{-j\pi \frac{\antdist}{\wavelength} (\numantennasper +1 - 2\antidx) \cos(\aod_{\useridx})}  ,
\end{align}
where $\aod_{\useridx}$ is the \gls{aod} from the satellite to user $\useridx$.
Under real circumstances, the estimate of the \glspl{aod} at the satellite might be flawed. We model this behavior as a uniformly distributed additive error ${\erroraod_{\useridx} \sim \mathcal{U}(-\errorbound, + \errorbound)}$ on the space angles $\cos(\aod_\useridx)$. In \cite{gracla2023satellite}, we show that this error can be interpreted as an overall multiplicative error ${\steeringvec_{\useridx}(\erroraod_{\useridx, \satidx})\in \mathbb{C}^{1 \times \numantennasper}} $ on our channel vector ${\csivector}_{\useridx}(\aod_{\useridx})$:

\begin{align}
\label{eq:error}
    \tilde{\csivector}_{\useridx}( \aod_{\useridx}, \erroraod_{\useridx}) = \csivector_{\useridx}( \aod_{\useridx}) \circ \steeringvec_{\useridx}\big(\erroraod_{\useridx}\big) \ .
\end{align}
In order to perform \gls{sdma}, we calculate a precoding vector $\precodingvec_\useridx \in \mathbb{C}^{\numantennasper \times 1}$ for each user $\useridx$ based on this estimate of the channel vector $\tilde{\csivector}_{\useridx}\in \mathbb{C}^{1 \times \numantennasper} $. Different approaches to determine the precoding vectors are discussed in the subsequent sections.
After calculating the precoding vector $\precodingvec_\useridx$, the data symbol $\datasymbol_\useridx$ of each user $\useridx$ is weighted with this precoding vector $\precodingvec_\useridx$. Taking complex \gls{awgn} $n_\useridx \sim \mathcal{CN}(0, \noisepower)$ into account, the received signal $y_\useridx$ for user $k$ results in

\begin{align}
\label{eq:transmission}
\textstyle
    y_\useridx = \csivector_\useridx\precodingvec_\useridx \datasymbol_\useridx + \csivector_\useridx\sum^\numusers_{\otheruseridx \neq \useridx}  \precodingvec_\otheruseridx \datasymbol_l + n_\useridx \ .
\end{align}
From \eqref{eq:transmission}, we get the \gls{sinr} $\Gamma_\useridx$ for user $\useridx$ 

\begin{align}
    \Gamma_\useridx = \frac{\left|\csivector_\useridx\precodingvec_{\useridx}\right|^2}{\noisepower+ \sum_{\otheruseridx \neq \useridx}^\numusers|\csivector_\useridx \precodingvec_{\otheruseridx}|^2} \ .
\end{align}
The achievable rate $\sumrate$, equal to the sum rate, is given by

\begin{align}
\label{eq:sumRate}
    \textstyle
    \sumrate = \sum_{\useridx = 1}^{\numusers} \log_2(1 + \Gamma_\useridx)
\end{align}
and serves as the performance metric for the different precoding approaches in this paper. In summary, our goal is to maximize the expected sum rate in presence of imperfect positional knowledge \eqref{eq:error} by learning a robust precoding algorithm using the \gls{sac} technique.

\begin{figure}[t]
	\begin{center}  
		\resizebox{0,54\linewidth}{!}{%
			\input{figures/verteilteSatellitenplus} 
		}
	\end{center}  
	\caption{%
        The single-satellite downlink scenario. The satellite is positioned at an altitude \( d_{0} \). Two users \( k, k-1 \), are positioned at \glspl{aod} \( \aod_{k}, \aod_{k-1} \). They are characterized by their channel vectors \( \csivector_{k}, \csivector_{k-1} \) and their inter-user distance~\( \userdist \).
    }
	\label{fig:geometry}  
\end{figure}

\section{Conventional Precoders}
\label{sec:conventionalprecoders}
This section introduces 1)~a conventional non-robust \gls{mmse} precoding approach and 2)~a robust \gls{slnr} precoder based on the second order statistics of the channel considering imperfect position knowledge. These analytical precoders serve as a benchmark for the learned precoders.

\subsection{MMSE}
The \gls{mmse} precoder is a well established precoder for scenarios with perfect channel state information~\cite{MMSEspace} and will serve as a baseline comparison in this work. For a channel estimation $\tilde{\csimatrix} = [\tilde{\csivector}_1 \dots \tilde{\csivector}_\numusers]^\text{T}$ the corresponding \gls{mmse} precoding matrix $\precodingmatrix^{\text{MMSE}} = [\precodingvec^\text{MMSE}_1 \dots \precodingvec_\numusers^\text{MMSE}]$ is given as 
\begin{align}
	\label{eq:MMSE}
	\begin{split}
		&\precodingmatrix^\text{MMSE}= \sqrt{\frac{\transmitpower}{\text{tr} \{ {{\precodingmatrix^{\prime}}^{\text{H}}} \precodingmatrix^{\prime} \} }} \cdot \precodingmatrix^{\prime}  
\\[0.8ex]
	&\precodingmatrix^{\prime} = \Big[ \mathbf{\tilde{\csimatrix}}^\text{H} \mathbf{\tilde{\csimatrix}} + \noisepower \frac{\numusers}{\transmitpower} \mathbf{I}_{\numantennasper} \Big]^{-1} \mathbf{\tilde{\csimatrix}}^\text{H} 
	\end{split} ,
\end{align}
where $\transmitpower$ denotes the overall transmit power of the satellite. We highlight that the \gls{mmse} approach is not always optimal in terms of the sum rate $\sumrate$ \eqref{eq:sumRate}.

\subsection{Robust SLNR}
The authors in~\cite{roper2023robust} have recently introduced an analytical robust precoding approach for a channel and error model equivalent to ours~\eqref{eq:error}. Because the optimization of the \glspl{sinr} is NP-hard \cite{roper2023robust}, the authors of \cite{roper2023robust} maximize with regard to the instantaneous \glspl{slnr} $\gamma_\useridx$ instead, with

\begin{align}
    \gamma_\useridx = \frac{\left|\csivector_\useridx\precodingvec_{\useridx}\right|^2}{\noisepower+ \sum_{\otheruseridx \neq \useridx}^\numusers|\csivector_\otheruseridx \precodingvec_{\useridx}|^2} \ .
\end{align}
Assuming equal power distribution among the users and considering the statistics of the estimation errors in the user positions, the optimization problem corresponds to maximizing the mean \gls{slnr} $\bar{\gamma}_\useridx=\mathbb{E}\{\gamma_\useridx\}$, i.e.,

\begin{align}
    \label{eq:optimizationproblemSLNR}
    \max_{\precodingvec_\useridx}\mathbb{E} \left\{ \frac{\left|\csivector_\useridx\precodingvec_{\useridx}\right|^2}{\noisepower+ \sum_{\otheruseridx \neq \useridx}^\numusers|\csivector_\otheruseridx \precodingvec_{\useridx}|^2} \right\} \quad \text{s.t.} \quad \precodingvec_\useridx^{\text{H}}\precodingvec_\useridx \leq \frac{\transmitpower}{\numusers} \ .
\end{align}
The precoding vector $\precodingvec_\useridx^\text{rSLNR}$ for each user $\useridx$, which satisfies \eqref{eq:optimizationproblemSLNR}, has been derived in \cite{roper2023robust} as 

\begin{align}
\label{eq:slnrprecoder}
    \precodingvec_\useridx^\text{rSLNR} = \sqrt{\frac{\transmitpower}{\numusers}} \psi_{\useridx, \max} \ ,
\end{align}
where $\psi_{\useridx, \max}$ is the eigenvector that corresponds to the largest eigenvalue of
\begin{align}
    \Big(\sum_{\otheruseridx \neq \useridx} \sigma_{\steeringvec_\otheruseridx}^2 \mathbf{R}_{\steeringvec_\otheruseridx} + \noisepower \frac{\numusers}{\transmitpower} \mathbf{I}_{\numantennasper}\Big)^{-1} \sigma_{\steeringentry_\useridx}^2 \mathbf{R}_{\steeringvec_\useridx} \ ,
\end{align}
with the inverse path loss $\sigma_{\steeringvec_\useridx}^2 = 1/\pathloss_\useridx$ and ${\mathbf{R}_{\steeringvec_\useridx} \in \mathbb{C}^{\numantennasper \times \numantennasper}}$ being the autocorrelation matrix of the steering vectors $\steeringvec_\useridx$.
The autocorrelation matrix of the steering vectors ${\mathbf{R}_{\steeringvec_\useridx}  = \mathbb{E}\{\steeringvec_\useridx \steeringvec_\useridx^{\text{H}}\}}$ is used because it has similar characteristics to the autocorrelation matrix of the channel $\mathbb{E}\{\csivector_\useridx \csivector_\useridx^{\text{H}}\}$, which is needed to solve \eqref{eq:optimizationproblemSLNR}. If we define an erroneous space angle $\hat{\phi}_\useridx = \cos(\aod_\useridx) + \erroraod_\useridx$ and use the definition of the steering vector from equation \eqref{eq:steering_vec}, we can rewrite the $[\antidx, \antidx']$-th element of $\mathbf{R}_{\steeringvec_\useridx}$ as

\begin{align}
    [\mathbf{R}_{\steeringvec_\useridx}]_{\antidx, \antidx'} &= \mathbb{E} \Big\{ \text{e}^{-j \frac{2\pi}{\wavelength}\antdist (\antidx - \antidx') (\hat{\phi}_\erroraod - \erroraod_\useridx)}   \Big\} \\
    &= \text{e}^{-j \frac{2\pi}{\wavelength}\antdist (\antidx - \antidx') \hat{\phi}_\erroraod} \varphi_\erroraod(\tfrac{2\pi}{\wavelength} \antdist (\antidx - \antidx')) \ ,
\end{align}
where $\varphi_\erroraod$ is the characteristic function that describes the probability distribution of the error. For a uniformly distributed error, $\varphi_\erroraod$ equals the sinc-function \cite{roper2023robust}

\begin{align}
    \varphi_\erroraod(t) = \text{sinc}(t\errorbound  ) \ .
\end{align}
Because the above precoder is optimized with regard to the mean \gls{slnr} and not the \gls{sinr}, it does not necessarily maximize the sum rate \eqref{eq:sumRate}. However, the authors in \cite{roper2023robust} show that, for perfect position knowledge and sufficiently large inter-user distances $\userdist$, the robust \gls{slnr} precoder is capacity achieving. We also note that this precoder always distributes the transmit power $\transmitpower$ evenly among the $\numusers$ users, even though cases with varying path losses between the users are probable.


%
	

\section{Reinforcement Learned Precoder}
\label{sec:OURAPPROACH}
Our goal in this section will be to find a function that takes as input the estimated channel matrix~\( \csimatrixestimate \) \refeq{eq:error} and outputs a precoding matrix $\precodingmatrix$ that maximizes the expectation~\( \bar{\sumrate} \) of the achievable sum rate~\( \sumrate \)~\refeq{eq:sumRate}. We will use a parameterized deep \gls{nn} \( \actornetworkvec_{\paramsactor} \), subsequently called \gls{acnn},  as a model function and then use the \gls{sac}~\cite{haarnoja2019soft} algorithm to tune this network's parameters~\( \paramsactor \).
\gls{sac} assumes the true system dynamics are unknown, \eg the distribution of errors on the estimated \gls{csit}~\( \csimatrixestimate \). Therefore, it trains a second \gls{nn} \( \criticnetworksca_{\paramscritic{}} \) that has parameters~\( \paramscritic{} \) in parallel to approximate the mapping of \( (\csimatrixestimate, \precodingmatrix) \rightarrow \bar{\sumrate} \). This known function \( \criticnetworksca_{\paramscritic{}} \), subsequently referred to as the \gls{crnn}, is used to as a guide to tune the precoding \gls{acnn}~\( \actornetworkvec_{\paramsactor} \). \gls{sac} recommends the use of multiple, independently initialized \gls{crnn} for stability~\cite{haarnoja2019soft}. In the following, we will omit the parameter indices \( \params \) in \gls{acnn}~\( \actornetworkvec_{\paramsactor}  \equiv  \actornetworkvec \) and \gls{crnn}~\( \criticnetworksca_{\paramscritic{}}  \equiv  \criticnetworksca \) for readability.

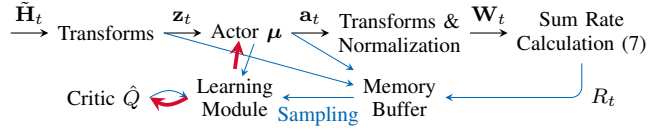
\begin{figure}[t]
    \centering
    \begin{tikzpicture}
\tikzstyle{A0} = [-, >={stealth}, rounded corners]
\tikzstyle{A1} = [->, >={stealth}, rounded corners]
\tikzstyle{A1learn} = [->, >={stealth}, rounded corners, color=uniblue2, line width=0.1mm]
\tikzstyle{A1paramupdate} = [->, >={stealth}, rounded corners, color=unired2, line width = 0.5mm]
\tikzstyle{A2} = [<->, >={stealth}, rounded corners]
\tikzstyle{selfloop} = [looseness=4]

\tikzstyle{every node}=[font=\footnotesize]

\newcommand{\xshift}{0.5}
\newcommand{\yshift}{0.1}

\node (origin)
    at (0.0, 0.0)
    []
    {};

\node (preprocessing)
    [right= \xshift of origin]
    {Transforms};

\node (actor)
    [right= \xshift of preprocessing]
    {Actor \( \actornetworkvec \)};

\node (reshapeprecoder)
    [right= \xshift of actor, align=center]
    {Transforms \& \\Normalization};

\node (commsystem)
    [right = \xshift of reshapeprecoder, align=center]
    {Sum Rate\\Calculation \refeq{eq:sumRate}};

\node (buffer)
    [below = \yshift of reshapeprecoder, align=center]
    {Memory\\Buffer};

\node (learningmodule)
    [left = 1.9*\xshift of buffer, align=center]
    {Learning\\Module};

\node (critic)
    [left = \xshift of learningmodule, align=center]
    {Critic \( \criticnetworksca \)};

\draw [A1]
    (origin)
        to node [above, align=center] {\( \tilde{\csimatrix}_{\timeindex} \)}
    (preprocessing);

\draw [A1]
    (preprocessing)
        to node [above, align=center] {\( \statevec_{\timeindex} \)}
    (actor);

\draw [A1]
    (actor)
        to node [above, align=center] {\( \actionvec_{\timeindex} \)}
    (reshapeprecoder);

\draw [A1]
    (reshapeprecoder)
        to node [above, align=center] {\( \precodingmatrix_{\timeindex} \)}
    (commsystem);

\draw [A1learn]
    (commsystem.south)
        |- node [right, align=center, color=black] {\( \sumrate_{\timeindex} \)}
    (buffer);

\draw [A1learn]
    (buffer)
        to
        node [below, align=center] {Sampling}
    (learningmodule);

\draw [A1learn]
    (preprocessing.east)
        to
    (buffer);

\draw [A1learn]
    (actor.east)
        to
    (buffer);

\draw [A1learn]
    ([xshift=0.1cm, yshift=0.1cm]actor.south)
        to
    ([xshift=0.1cm, yshift=-0.1cm]learningmodule.north);

\draw [A1learn]
    (critic.east)
        to [bend left, looseness=1.2]
    (learningmodule.west);

\draw [A1paramupdate]
    (learningmodule.north)
        to 
    ([xshift=-0.1cm, yshift=-0.1cm]actor);

\draw [A1paramupdate]
    (learningmodule.west)
        to [bend left, looseness=1.2]
    (critic.east);

\end{tikzpicture}





    \caption{The learned \gls{sac} precoders' process flow. Black arrows form an inference step, blue arrows show the components that are collected for a learning step, and red arrows show \gls{nn} parameter updates.}
    \label{fig:sac_flowchart}
\end{figure}

Learning to optimize the \gls{nn} parameters via \gls{sac} is comprised of two independent components: 1)~data generation; 2)~parameter updates. \reffig{fig:sac_flowchart} gives an overview of the process flow, we describe it in more detail in the following.
First, to generate the data to learn from, we perform an inference step~\( \timeindex \), starting by obtaining a complex-valued channel state estimate~\( \csimatrixestimate_{\timeindex} \in \mathbb{C}^{\numusers\times \numantennasper } \). We then flatten it into a vector and transform it to a real vector~\( \tilde{\statevec} \in \mathbb{R}^{1 \times 2\numusers \numantennasper} \) as it is easier for \gls{nn} to digest. For the real-complex transformation we consider decomposition into a)~real and imaginary part, as well as b)~decomposition into magnitude and phase, the choice of which we will discuss briefly later.
The inputs~\( \tilde{\statevec} \) are also standardized to approximately zero mean and unit variance using static scaling factors, which is known to promote convergence speed in \gls{nn}~\cite{goodfellow2017deep}. We discuss this choice in \refsec{sec:implementation}.
We forward the standardized vector~\( {\statevec_{\timeindex} \in \numbersreal^{\num{1} \times 2\numusers \numantennasper} }\) through the \gls{acnn}, which is a standard feed-forward \gls{nn} with four times as many outputs as we require entries for a precoding matrix~\( \precodingmatrix \). The outputs are grouped in pairs of two, where one of each pair represents the mean and the other the scale of a Normal distribution to sample from. This formulation gives us a measure of uncertainty about each output that we will make use of during the training step.
After sampling from the distributions given by the output pairs, we transform this output~\( {\actionvec \in \numbersreal^{\num{1} \times 2\numusers \numantennasper}} \) into a complex vector of half length and reshape to a precoding matrix.
Finally, we rescale the matrix to the available signal power and gain a normalized precoding matrix~\( \precodingmatrix_{\timeindex}  \in \mathbb{C}^{ \numantennasper \times \numusers}\).
In testing, we find that magnitude-phase decomposition works best for the network input and real-imaginary composition works best for the output, though this is still under investigation.
To conclude a data generation step, we evaluate the sum rate~\( \sumrate_{\timeindex} \) achieved by this precoding matrix~\( \precodingmatrix_{\timeindex} \) and store the tuple of \( (\statevec, \actionvec, \sumrate)_{\timeindex} \) in a memory buffer for the learning step.

Learning steps are performed using the \gls{sgd} principle of noisily approximating a gradient step based on a batch subset of the entire data set. During a learning step, first, a batch \( \batchset = \{ (\cdot)_\batchindex~|~\batchindex \sim \mathcal{U}(\mathbf{0}, \mathbf{\expbuffersize}) \} \) of \( |\batchset| = \batchsize \) tuples are drawn from the memory buffer of size~\( \expbuffersize \). Next, the \glspl{crnn}~\( \criticnetworksca \) are updated. A loss is calculated as follows,
\begin{align}
\label{eq:losscritic}
    \losscritic =
        \frac{1}{\batchsize}
        \sum_{\batchindex \in \batchset}
            (
                \criticnetworksca(\statevec_{\batchindex}, \actionvec_{\batchindex})
                -
                \sumrate_{\batchindex}
            )^{2}
            +
            \weightregularizationcriticscale \| \paramscritic{} \|^{2}
    ,
\end{align}
where the first term describes the mean square error in sum rate estimation and the second term is a weight regularization that is discussed later. \( \weightregularizationcriticscale \)~is a scaling term. A \gls{sgd}-like update is performed to minimize this batch loss.
Next, the \gls{acnn} is updated. Its loss contains three sum terms:
\begin{align}
\label{eq:lossactor}
    \lossactor = 
        &\frac{1}{\batchsize}
        \sum_{\batchindex \in \batchset}
            -\criticnetworksca(\statevec_{\batchindex}, \actornetworkvec(\statevec_{\batchindex})) \\
        +& \frac{1}{\batchsize}
            \sum_{\batchindex \in \batchset}
                \exp (\logentropyscale) \log(\pi(\actornetworkvec(\statevec_{\batchindex}))) \\
        +& \weightregularizationactorscale \| \paramsactor \|
    ,
\end{align}
where \( \logentropyscale, \weightregularizationactorscale \) are scaling terms.  Recalling that the precoder is sampled from a distribution parameterized by the outputs of the \gls{acnn}, \( \pi(\cdot) \) is the probability of the sampled precoding, given this distribution. The first term calls the optimization to maximize the estimated sum rate.
The second term encourages the optimization to increase the output variance where it does not foresee gains in sum rate from keeping the variance tight.
It thereby encourages the \gls{acnn} to explore more thoroughly where no good solution has been found yet.
The third term is, again, a weight regularization. If multiple \gls{crnn} are used, we take the minimum sum rate estimate per tuple for a conservative estimate.

In the following section, we discuss specific implementation choices such as the weight regularization, and then proceed to evaluate the learned precoder and the two benchmarks.


%
	

\section{Evaluation}
\label{sec:experiments}
Here, we discuss specific implementation details that we found crucial in learning a precoder. We then present performance comparisons and discuss the relative advantages and disadvantages of the precoders under study. The full code implementation, the trained models, their training configurations and supplemental figures are available online~\cite{github}. \reftab{tab:parameters} lists a selection of important system and learning parameters.

\subsection{Implementation Details}
\label{sec:implementation}

\begin{table}[!t]
	\renewcommand{\arraystretch}{1.3}
	\caption{Selected Parameters}
	\label{tab:parameters}
	\centering
    \addtolength{\tabcolsep}{-.06cm}
	\rowcolors{1}{white}{uniblue1!5} 
	\begin{tabular}{llll}
		\hline
        Noise Power $\noisepower$ & $6\text{e-}13\,\text{W}$
            &
        Transmit Power $\transmitpower$ & \SI{100}{W}
        \\
        Satellite Altitude $d_0$ & \SI{600}{\km} 
            &
        Antenna Nr. $\numantennasper$ & \{\num{10},~\num{16}\}
        \\
         User Nr. $\numusers$ & \num{3}
			&
		Overall Sat. Gain & \SI{20}{dBi} 
        \\
        Wavelength $\wavelength$ & \SI{15}{cm}
			&
		Gain per User $\usergain$ & \SI{0}{dBi}
		\\
        Inter-Ant.-Distance $\antdist $ & $ 3 \wavelength/2$
            &
        \gls{sgd} Optimizer & Adam
        \\
        Training batch size \( \batchsize \) & \( \num{1024} \)
            &
        Init. LR \gls{crnn}, \gls{acnn} & \( 1\text{e-}4, 1\text{e-}5 \)
        \\
        Learning Buffer Size & \num{100000} 
            &
        Inference / Learning & \( 10:1 \)
        \\
        L2 Scales \( \weightregularizationcriticscale, \weightregularizationactorscale \)    & \num{0.1}
            &
        log Entropy Scale \( \logentropyscale \) & Var.
        \\
		\hline
	\end{tabular}
\end{table}

As discussed in the previous section, we find rescaling the network inputs to be highly beneficial for fast and stable learning, in accordance with theoretical literature~\cite{goodfellow2017deep}. It could be argued that taking a large number of samples to find the population means and scales before starting to learn is sample inefficient and undesirable, however, doing a hypothesis test at a significance level of~\( \SI{5}{\percent} \), we find the statistics of the population to be approximable within \( \pm\SI{10}{\percent} \) by taking just \( \num{100} \) samples. For similar reasons, rescaling values not just at the input but also between network layers is desirable and proves beneficial, with intermediate Batch Normalization layers~\cite{balestriero_batch_2022} being the most common approach. Our \gls{nn} implementations correspondingly use four fully connected layers (\( \num{512} \) nodes each) stacked alternatingly with Batch Normalization layers.

Another critical change to our prior work is the addition of weight regularization terms in \refeq{eq:losscritic}, \refeq{eq:lossactor}. Weight regularization is another standard method, the reasons for its positive influence still being under intense scrutiny~\cite{andriushchenko2023weightdecay}. We find weight regularization to be specifically favorable for training in the presence of error, which intuitively can be interpreted as stopping a learning step from committing overmuch to a single batch of channel realizations.
In a similar vein, we significantly increase the number of experiences held in the memory buffer compared to our earlier work (\( 10\text{k} \rightarrow 100\text{k} \)) to provide a more rich set of training experiences to sample from. To compensate, we also adjust the ratio of inference steps per learning step~(\( 1:1 \rightarrow 10:1 \)) so that the age of the oldest experience in the buffer stays the same, as per~\cite{fedus_revisiting_2020}.

Overall, we find that the most significant parameter choice is the \gls{sgd} learning rate, which we keep variable with a Cosine Decay schedule in the area of \( \num{1}\text{e-}\num{4}, \num{1}\text{e-}\num{5} \) for \gls{crnn}, \gls{acnn} respectively. For the detailed configuration we again refer to the repository~\cite{github}. We also highlight that, considering practicality, hyper parameter search and training time are limited, thus, our trained models certainly do not present the optimum solution.
Training is performed until no significant performance increase is observed, up to around \( \num{1}\text{e}\num{6} \) to \( \num{14}\text{e}\num{6} \) simulation steps~\( \timeindex \) depending on the scenario.

\subsection{Evaluation Design}
Each simulation will assume a certain mean user distance~\( \meanuserdist \). For each simulation step~\( \timeindex\), each user will be assigned a new position uniform randomly within~\( \pm \meanuserdist/2 \) around their mean position, and channel conditions will be updated according to \refsec{sec:setup}.
We investigate the following three scenarios: a) $\numantennasper = 10$ satellite antennas, mean user distance~\( \meanuserdist = \SI{100}{\km} \); b)~$\numantennasper = 16$,~\( \meanuserdist = \SI{100}{\km} \); c)~$\numantennasper = 16$,~\( \meanuserdist = \SI{10}{\km} \).
Before evaluation, we train two learned precoders on each scenario: 1)~trained with perfect \gls{csit}, marked with blue square \eg~{\psac{1}}; 2)~trained at error bound~\(\errorbound = \num{0.05} \), marked with green cross, \eg~{\rsac{1}}.
Evaluations are repeated with \num{1000} Monte Carlo iterations to account for the stochastic elements of the simulation design, we evaluate the mean performance and its standard deviation.

\subsection{Results}
\label{sec:results}
We first explore scenario~a), where a mean user distance \( \meanuserdist = \SI{100}{\km} \) ensures, on average, good spatial partitioning of the users and \(\numantennasper = \num{10}\) satellite antennas will produce wide beams relative to user distances. We train two precoders, \psac{1} with perfect \gls{csit} and \rsac{1} with erroneous \gls{csit}, and then compare their mean performance when evaluated on increasingly unreliable \gls{csit}. \reffig{fig:results_10ant} presents the mean performance at increasingly large settings of error bound~\( \errorbound \) for the two trained schedulers as well as the \gls{mmse} and robust \rslnr precoders from \refsec{sec:conventionalprecoders}.
As expected, all precoders' performances are impacted by increasingly unreliable \gls{csit}. Precoders \gls{mmse}, robust \rslnr and \psac{1} achieve comparable performance for perfect \gls{csit}~(\( \errorbound = \num{0.0} \)), while \rsac{1}, having only encountered unreliable \gls{csit} during training, has adopted a strategy that does not scale as well with reliable information. On the other hand, \rsac{1} shows the least performance degradation as the \gls{csit} becomes increasingly unreliable. It takes the performance lead at its training point of \( \errorbound = \num{0.05} \), with the performance gap increasing thereafter. The \gls{mmse} precoder, expectedly, shows the worst performance with unreliable \gls{csit}, while the robust \rslnr and \psac{1} precoders are closely matched. This result might, however, be slightly misleading, as robust \rslnr and \psac{1} have adopted significantly different strategies, which we will see in the following.

\begin{figure}[!t]
    \centering
    \input{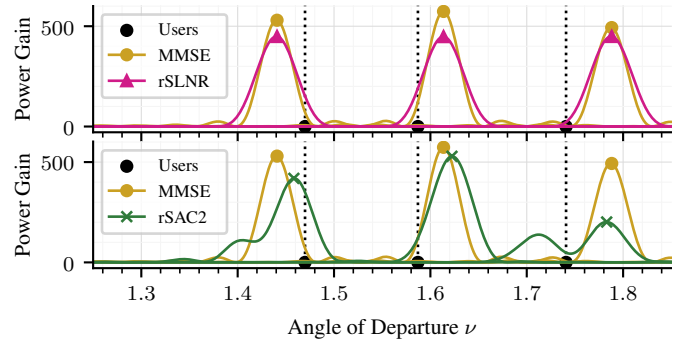}
    \caption{%
        Beam patterns for one specific simulation realization of scenario~b). Curves represent the precoding vectors of the different users at different \glspl{aod}~\( \aod \) in linear scale. Sum rates~\( \sumrate \) achieved: \gls{mmse}:~\SI{0.86}{\bit/\second/\Hz}, \rslnr:~\SI{1.54}{\bit/\second/\Hz}, \rsac{2}:~\SI{2.07}{\bit/\second/\Hz}.
    }
    \label{fig:beampattern_16ant_sac2}
\end{figure}

We repeat this experiment for scenario~b), where the increased number of $N=\num{16}$ satellite antennas allows for more narrow beams. This enables better user separation even at close user distances, at the cost of more severe performance drops when a beam is missteered. The results are displayed in \reffig{fig:results_16ant} and follow the same trajectory as scenario a), though we see that the narrower beams lead to a more pronounced performance drop as the unreliability~\( \errorbound \) increases.
In order to understand how each precoder achieves their robustness, we take a look at their beam patterns for a specific simulation realization. \reffig{fig:beampattern_16ant_sac1} compares the beam patterns of the \rslnr and our learned precoder \psac{2} with the \gls{mmse} precoder. We select this plot as a representative for the precoders' behavior, though we highlight that it depicts just one realization of user positions, error values, large scale fading. Further beam patterns are provided in~\cite{github}. Black dots and dotted lines represent the true user positions, whereas the erroneously estimated positions can be discerned by the placement of the \gls{mmse} precoder's beams. We observe in the top figure that, compared to the \gls{mmse} precoder, the robust \rslnr precoder achieves its robustness by trading beam height in favor of beam width, covering a larger area. On the contrary, the \psac{3} precoder in the bottom figure, not having encountered unreliable information during training, has no incentive to opt for wider beams. Nevertheless, we observe that it achieves robustness by two other factors: 1)~power allocation among the different user's beams; 2)~better user tracking. We report that it achieves the second factor by exploiting the power fading information of the \gls{csit}. In the depicted realization, the center user's large scale fading is near one, hence, the power fading of the \gls{csit} closely corresponds to a certain satellite-to-user distance that the learned precoder uses to fine-tune its beam positioning. In \reffig{fig:beampattern_16ant_sac2}, we repeat this comparison for the \rsac{2} precoder that was trained for robustness to severe errors. We observe that this scheduler makes use of more irregular beam shapes to cover wide areas.

\begin{figure}[!t]
    \centering
    \input{figures/error_sweep_10_ant_paper.pgf}
    \caption{Scenario~a), \( \numantennasper = \num{10} \) satellite antennas, \( \meanuserdist = \SI{100}{\km} \) mean user distance. Testing precoders mean performance with increasing error bounds~\( \errorbound \). \psac{1} is trained with perfect \gls{csit}, \rsac{1} is trained at \( \errorbound = \num{0.05} \). Markers on the horizontal axis show the error bound that \psac{1} resp. \rsac{1} were trained at.}
    \label{fig:results_10ant}
\end{figure}

\begin{figure}[!t]
    \centering
    \input{figures/error_sweep_16_ant_paper.pgf}
    \caption{Scenario~b), \( \numantennasper = \num{16} \) satellite antennas, \( \meanuserdist = \SI{100}{\km} \) mean user distance. Testing precoders mean performance with increasing error bounds~\( \errorbound \). \psac{2} is trained with perfect \gls{csit}, \rsac{2} is trained at \( \errorbound = \num{0.05} \). Markers on the horizontal axis show the error bound that \psac{2} resp. \rsac{2} were trained at.}
    \label{fig:results_16ant}
\end{figure}

\begin{figure}[!t]
    \centering
    \input{figures/beampattern_16ant_learned_no_error.pgf}
    \caption{%
        Beam patterns for one specific simulation realization of scenario~b). Curves represent the precoding vectors of the different users at different \glspl{aod}~\( \aod \) in linear scale. Sum rates~\( \sumrate \) achieved: \gls{mmse}:~\SI{0.83}{\bit/\second/\Hz}, \rslnr:~\SI{1.33}{\bit/\second/\Hz}, \psac{2}:~\SI{2.15}{\bit/\second/\Hz}.
    }
    \label{fig:beampattern_16ant_sac1}
\end{figure}

Finally, in scenario~c), we study a case in which the \gls{mmse} and \rslnr precoders are not sum rate optimal due to very close average user positioning ($\meanuserdist = \SI{10}{\km}$) relative to the beam width. \reffig{fig:results_16ant_close} displays again a performance sweep over increasingly unreliable \gls{csit} for this scenario. We see the learned precoders \psac{3}, \rsac{3} attain greater sum rate performances, achieved by simply not allocating any power to the center user such that it does not interfere with the other two users. We also see that the robust \rslnr precoder, with its preassumptions violated by the high spatial coupling of user channels, is not able to provide robustness in this scenario even compared to the \gls{mmse} precoder.
The learned precoders achieve very high degrees of robustness, though we must qualify this result stemming from the erroneous channel realizations being much larger than the variation through user positioning in this scenario. The learned precoders exploit this by discarding implausible \gls{csit}, which we do not expect to be possible to the same degree under real circumstances.

In summary, learned precoders offer high flexibility, adjusting to various combinations of user constellations, satellite configurations and error influences, while achieving strong performance. Analytical precoders, while perhaps more predictable, can be complex and costly to solve mathematically and may suffer greatly when the real conditions do not match those that the precoder was modeled on.

\begin{figure}[!t]
    \centering
    \input{figures/error_sweep_16_ant_close_paper.pgf}
    \caption{Scenario~c), \( \numantennasper = \num{16} \) satellite antennas, \( \meanuserdist = \SI{10}{\km} \) mean user distance. Bad conditions for \gls{mmse} and \rslnr precoders. Testing precoders mean performance with increasing error bounds~\( \errorbound \) . \psac{3} is trained with perfect \gls{csit}, \rsac{3} is trained at \( \errorbound = \num{0.05} \). Markers on the horizontal axis show the error bound that \psac{3} resp. \rsac{3} were trained at.}
    \label{fig:results_16ant_close}
\end{figure}



%
	

\section{Conclusions}
\label{sec:conclusions}

In this paper we studied the influence of imperfect user position knowledge on \gls{sdma} precoding in \gls{leo} satellite downlink scenarios. Our goal was to design a robust precoding approach that manages inter-user interference for arbitrary error sources and channels. 
Using data-driven deep \gls{rl} via the \gls{sac} algorithm, we built learned precoders that obtained high performance in terms of both achievable rate and robustness to positioning errors. We qualified these results in comparison to two analytical benchmark precoders, the conventional \gls{mmse} precoder as well as a robust precoder that leverages stochastic channel information.
Their flexibility at high performance could make learned satellite precoding algorithms an attractive candidate for use in 6G and beyond.


%
	%
	%
	%
	\bibliographystyle{IEEEtran}%
	{%
		\makeatletter  
			\clubpenalty=10000  
			\@clubpenalty=\clubpenalty
			\widowpenalty=10000
		\makeatother
		\bibliography{%
			ref/IEEEabrv,
			ref/references
		}%
	}%
	%
	%
	%
\end{document}